\documentclass[twocolumn,preprintnumbers,amsmath,amssymb]{revtex4}

\usepackage{array}
\usepackage{booktabs}
\usepackage{tabu}
\usepackage{dcolumn}
\usepackage{amsmath}
\usepackage{amsfonts}
\usepackage{amssymb}
\usepackage{graphicx}
\usepackage{subfigure}
\usepackage{graphicx}
\usepackage{dcolumn}
\usepackage{bm}
\usepackage{xcolor}
\def\be{\begin{equation}}
  \def\ee{\end{equation}}
\def\bea{\begin{eqnarray}}
\def\eea{\end{eqnarray}}
\def\f{\frac}
\def\n{\nonumber}
\def\l{\label}
\def\p{\phi}
\def\o{\over}
\def\R{\rho}
\def\pa{\partial}
\def\om{\omega}
\def\na{\nabla}
\def\P{\Phi}

\begin{document}

\title{Tightening  the entropic uncertainty relations for multiple measurements and applying it to quantum coherence}

\author{H. Dolatkhah}
\affiliation{Department of Physics, University of Kurdistan, P.O.Box 66177-15175, Sanandaj, Iran}
\author{S. Haseli}
\affiliation{Department of Physics, Urmia University of Technology, Urmia, Iran}
\author{S. Salimi}
\email{shsalimi@uok.ac.ir}
\author{A. S. Khorashad}
\affiliation{
Department of Physics, University of Kurdistan, P.O.Box 66177-15175, Sanandaj, Iran.\\}
\date{\today}

\def\be{\begin{equation}}
  \def\ee{\end{equation}}
\def\bea{\begin{eqnarray}}
\def\eea{\end{eqnarray}}
\def\f{\frac}
\def\n{\nonumber}
\def\l{\label}
\def\p{\phi}
\def\o{\over}
\def\R{\rho}
\def\pa{\partial}
\def\om{\omega}
\def\na{\nabla}
\def\P{\Phi}

\begin{abstract}
{\bf The uncertainty principle sets limit on our ability to predict the values of two incompatible observables measured on a quantum particle simultaneously. This principle can be stated in various forms. In quantum information theory, it is  expressed in terms of the entropic measures. Uncertainty bound can be altered by considering a particle as a quantum memory correlating with the primary particle. In this work, we provide a method for converting the entropic uncertainty relation in the absence of quantum memory to that in its presence. It is shown that the lower bounds obtained through the method are tighter than those having been achieved so far. The method is also used to obtain the uncertainty relations for multiple measurements in the presence of quantum memory. Also for a given state, the lower bounds on the sum of the relative entropies of unilateral coherences are provided using the uncertainty relations in the presence of quantum memory, and it is shown which one is tighter.}

\end{abstract}
\maketitle

\section{INTRODUCTION}	
{\bf The uncertainty principle is the central core of quantum theory. It represents the unpredictability of quantum phenomena. The principle sets limits on the precise prediction of the outcomes of two incompatible quantum measurements on
a particle \cite{Heisenberg}. The uncertainty principle can be stated in various forms. The most famous form of the uncertainty relation was presented by Robertson \cite{Robertson}  and Schr\"{o}dinger \cite{Schrodinger}. They showed that for arbitrary pairs of noncommuting observables  $Q$ and $R$, the uncertainty relation has the following form,}
\begin{equation}\label{Roberteq}
\Delta Q \Delta R \geq \frac{1}{2}\vert \langle \left[ Q, R \right] \rangle \vert,
\end{equation}
where $\Delta Q(\Delta R)$ indicates the standard deviation of the associated observable $ Q (R)$,
\begin{equation}\label{deviation}
\Delta Q = \sqrt{\langle Q^{2}\rangle -\langle Q \rangle^{2}}  \quad (\Delta R = \sqrt{\langle R^{2}\rangle -\langle R \rangle^{2}}).
\end{equation}
{\bf This form of the uncertainty relation is still one of the best well-known ones. Utilizing the Shannon entropy as an appropriate measure for the uncertainty is a more efficient way to represent the uncertainty relation which was conjectured by Deutsch \cite{Deutsch} who introduced the entropic uncertainty relation. Deutsch's inequality was improved by Kraus \cite{Kraus}, and proved by Massen and Uffink later \cite{Uffink}. This uncertainty relation states that for two observables $X$ and $Z$ with eigenbases $\lbrace \vert x_i\rangle \rbrace$ and $\lbrace \vert z_i\rangle \rbrace$, respectively, and for any state $\rho$, one can write }
\begin{equation}\label{Maassen and Uffink}
H(X)+H(Z)\geq \log_2 \frac{1}{c}\equiv q_{MU},
\end{equation}
where $H(O) = -\sum_{k} p_k \log_2 p_k$, is the Shannon entropy of the measured observable $O \in \lbrace X, Z \rbrace$, $p_k$ is the probability of the outcome $k$, {\bf the quantity $c$ is defined as $c = max_{i,j} c_{ij}$, in which} $c_{ij} = \vert\langle x_i \vert z_j\rangle \vert ^{2}$, and $q_{MU}$ is called incompatibility measure. \\
{\bf There are many applications of the uncertainty relation in the field of quantum information:} quantum key distribution \cite{Koashi,Berta}, quantum random number generation \cite{Vallone,Cao}, entanglement witness \cite{Berta2}, EPR steering \cite{Walborn,Schneeloch}, and quantum metrology \cite{Giovannetti}. {\bf A huge amount of effort has been made to expand and modify this relation \cite{Berta,Coles1,Bialynicki,Pati,Ballester,Vi,Wu,Wehner,Rudnicki,Rudnicki1,Pramanik,Maccone,Pramanik1, Zozor,Coles,Adabi,Adabi1,Liu,Kamil,Zhang,R}. Berta \emph{et al.} studied it in the presence of the quantum memory \cite{Berta}.

Let us see how one can generalize  the entropic uncertainty relation to the one describing a situation in which an extra quantum system as the quantum memory $B$ correlates with the measured quantum system $A$. It can be done by an interesting game between two observers, Alice and Bob. At the beginning of the game, Bob prepares a particle in an arbitrary quantum state and sends it to Alice, then both of them reach an agreement about measuring two observables $X$ and $Z$ by Alice on the particle. Alice performs her measurement on the quantum state of the particle, and announces her choice of the measurement to Bob. Bob seeks to minimize his uncertainty about Alice's measurement outcome; its minimum is bounded by Eq.\;(\ref{Maassen and Uffink}). As mentioned before, there is only one particle in the above process. If a correlated bipartite state $\rho_{AB}$ is prepared by Bob, and only one of them is sent to Alice and the other one is kept as the quantum memory, Bob is able to guess Alice's measurement outcomes with better accuracy. The uncertainty relation in the presence of the quantum memory is expressed as \cite{Berta}  }
\begin{equation}\label{Berta}
H(X \vert B)+H(Z \vert B) \geq q_{MU} +S(A \vert B),
\end{equation}
where $H(X \vert B) = S(\rho_{XB})-S(\rho_{B})$ and $H(Z \vert B) = S(\rho_{ZB})-S(\rho_{B})$ are the conditional von Neumann entropies of the post measurement states
\begin{eqnarray*}
  \rho_{XB} &=& \sum_{i}(\vert x_{i}\rangle\langle x_{i}\vert\otimes I ) \rho_{AB}(\vert x_{i}\rangle\langle x_{i}\vert\otimes I ),\\
  \rho_{ZB} &=& \sum_{j}(\vert z_{j}\rangle\langle z_{j}\vert\otimes I ) \rho_{AB}(\vert z_{j}\rangle\langle z_{j}\vert\otimes I ),
\end{eqnarray*}
and $S(A|B) = S(AB)- S(B)$ is the conditional von Neumann entropy. {\bf There exist three special cases:} firstly, if the measured part $A$ and the quantum memory $B$ are entangled, the uncertainty lower bound (ULB) reduces due to the negativity of the conditional entropy $S(A \vert B)$, {\bf and Bob's uncertainty about Alice's measurement outcomes can be reduced.} Secondly, {\bf when $A$ and $B$ are maximally entangled, one obtains $S(A \vert B)=-\log_{2}d$, where $d$ is known as the dimension of the measured particle. Since $\log_{2}{1 \over c}$ cannot exceed $\log_{2}d$, Bob can exactly guess Alice's measurement outcomes. Finally,} in the absence of the quantum memory, Eq.\;(\ref{Berta}) reduces to
\begin{equation}\label{Berta2}
H(X)+H(Z)\geq q_{MU}+S(A),
\end{equation}
which is tighter than Eq.\;(\ref{Maassen and Uffink}), due to this fact that the measured particle is in a mixed state {\bf which leads to} $S(A)>0$. {\bf Pati \emph{et al.} proved that the uncertainties are lower bounded by a term added to the right hand side of Eq.\;(\ref{Berta}) \cite{Pati}}
\begin{align}\label{pati}
H(X \vert B)+H(Z \vert B) & \geq q_{MU} + S(A \vert B) \\ \nonumber
& + \max\lbrace 0,D_{A}(\rho_{AB})-J_{A}(\rho_{AB}) \rbrace,
\end{align}
{\bf where $J_{A}(\rho_{AB})$ is the classical correlation and defined as
\begin{equation}\label{classical}
J_{A}(\rho_{AB})=S(\rho_{B})-\min_{\lbrace \Pi_{i}^{A} \rbrace}S(\rho_{B\vert \lbrace \Pi_{i}^{A} \rbrace}),
\end{equation}
in which minimization is performed over all positive operator-valued measures (POVMs)} ${\lbrace \Pi_{i}^{A} \rbrace}$ acting on the measured part $A$.  $D_{A}(\rho_{AB})$ is  called quantum discord and defined as the difference between the total correlation $I(A:B)=S(\rho_{A})+S(\rho_{B})-S(\rho_{AB})$ and the classical one. {\bf Note that Pati's lower bound is tighter than that of Berta if the discord $D_{A}(\rho_{AB})$ is larger than the classical correlation $J_{A}(\rho_{AB})$. Adabi \emph{et al.} showed that the uncertainties $H(X \vert B)$ and $H(Z\vert B)$ are lower bounded by adding a term to the right hand side of Eq.\;(\ref{Berta}) \cite{Adabi}. The term depends on the Holevo quantity and mutual information,}
\begin{equation}\label{new1}
H(X \vert B)+H(Z \vert B)\geq q_{MU} + S(A|B)+\max\{0 , \delta\},
\end{equation}
where  $$\delta=I(A:B)-[I(X:B)+I(Z:B)],$$
and
$$I(P:B)= S(\rho_{B})- \sum_{i}p_{i}S(\rho_{B|i})$$
is the Holevo quantity and is equal to the upper bound of the accessible information to Bob about Alice's measurement outcomes. Note that when observable $P$ is measured  on the  part A, the $i$-th outcome with probability $p_{i}= Tr_{AB}(\Pi^{A}_{i}\rho_{AB}\Pi^{A}_{i})$ is obtained and the part $B$ is left in the corresponding state $\rho_{B|i}= \frac{Tr_{A}(\Pi^{A}_{i}\rho_{AB}\Pi^{A}_{i})}{p_{i}}$. It is worth noting that this lower bound is tighter than both the Berta and Pati lower bounds. If the part $B$ is removed, Eq.\;(\ref{new1}) reduces to:
\begin{equation}\label{Maseen}
H(X)+H(Y)\geq q_{MU}+S(A),
\end{equation}
which is the same as Eq.\;(\ref{Berta2}). The main purpose of this paper is to provide a method which can be applied to convert the uncertainty relations in the absence of quantum memory to those in the presence of it. It is also shown that the unilateral coherence of a given bipartite quantum system in one measurement basis is restricted by the unilateral coherence of the same quantum system in other measurement basis. {\bf As a result, quantum uncertainty relations can be written in terms of unilateral coherence}. The paper is organized as follows: In Sec. \ref{Sec2} a general method is proposed to convert uncertainty relations to those in the presence of quantum memory. In Sec. \ref{Sec3} the method is applied to entropic uncertainty relations for multiple measurements. Sec. \ref{Sec4} presents a brief review on quantum coherence. {\bf In Sec. \ref{Sec5}, it is shown that how one can obtain the uncertainty relations for unilateral coherence by using  the uncertainty relations in the presence of quantum memory}. Finally, the results are summarized in Sec. \ref{conclusion}.

\section{{\bf A method for obtaining the uncertainty relations in the presence of quantum memory}}
\label{Sec2}
In this section, a method is introduced by which one can convert the uncertainty relations in the absence of quantum memory to those in the presence of it.  Assume that the general form of the uncertainty relation for $N$ measurements $M_{1},M_{2},$ ...,$M_{N}$ is
\begin{equation}\label{general}
\sum^{N}_{m=1} H(M_{m})\geq LB,
\end{equation}
where $LB$ is an abbreviation for lower bound. Using the Holevo quantity, this relation can be transformed into the uncertainty relation in the presence of quantum memory.

{\bf To achieve this aim, one can use the definition of the von Neumann conditional entropy, $H(M_{m} \vert B) = S(\rho_{M_{m}B})-S(\rho_{B}),$ and that of the mutual information, $I(M_{m}:B)=H(M_{m})+S(\rho_B)-S(\rho_{M_{m}B})$. Adding the two quantities, one obtains \cite{Nielsen}
\begin{equation}\label{identity}
H(M_{m})=H(M_{m}\vert B)+I(M_{m}:B).
\end{equation}
Subtracting both sides of Eq.\;(\ref{general}) by $\sum^{N}_{m=1}I(M_{m}:B)$ and using Eq.\;(\ref{identity}), one arrives at }
\begin{equation}\label{general 2}
\sum^{N}_{m=1}H(M_{m}\vert B)\geq LB-\sum^{N}_{m=1}I(M_{m}:B),
\end{equation}
which is an uncertainty relation in the presence of quantum memory. As can be seen, {\bf it is a simple way which can be used to convert an entropy-based uncertainty relation in the absence of quantum memory to that in its presence.}

\section{Entropic uncertainty relations for multiple measurements}\label{Sec3}
Several entropic uncertainty relations for multiple measurements have been proposed \cite{Liu,Zhang,Yunlong}. 
{\bf Recently, Liu \emph{et al.}  derived an entropic uncertainty relation for $N$ measurements $M_{m}$ as \cite{Liu}
\begin{equation}\label{Liu}
\sum^{N}_{m=1} H(M_{m})\geq -\log_{2}(b)+(N-1)S(\rho),
\end{equation}
in which $$
 b=\max_{i_{N}} \left\{ \sum_{i_{2}\sim{i_{N-1}}} \max_{i_{1}} \Big[ |\langle u^{1}_{i_{1}}|u^{2}_{i_{2}}\rangle |^{2} \Big]\prod^{N-1}_{m=2}|\langle u^{m}_{i_{m}}|u^{m+1}_{i_{m+1}}\rangle |^{2} \right\} ,$$
where $|u^{m}_{i_{m}}\rangle$ is the $i$-th eigenvector of $M_{m}$, and $ S(\rho)=-Tr(\rho\log_2 \rho) $ is the von Neumann entropy of quantum state $ \rho $.
In the presence of quantum memory the relation is converted to:
\begin{equation}\label{Liu 2}
\sum^{N}_{m=1}H(M_{m}\vert B)\geq -\log_{2}(b)+(N-1)S(A\vert B),
\end{equation}
where b is the same as that in Eq.\;(\ref{Liu}).
{\bf Substituting Eq.\;(\ref{identity}) into Eq.\;(\ref{Liu}), one obtains }
\begin{equation}\label{Liu 33}
\sum^{N}_{m=1}H(M_{m}\vert B)\geq -\log_{2}(b)+(N-1)S(\rho)-\sum^{N}_{m=1}I(M_{m}:B).
\end{equation}
Using $ S(\rho)=S(A\vert B)+I(A:B) $ in Eq.\;(\ref{Liu 33}), one comes to
\begin{eqnarray}\label{Liu 3}
\sum^{N}_{m=1}H(M_{m}\vert B) & \geq & -\log_{2}(b)+(N-1)S(A\vert B)  \\
    &   & +(N-1)I(A:B) -\sum^{N}_{m=1}I(M_{m}:B), \nonumber
\end{eqnarray}
which can be rewritten as
\begin{equation}\label{Liu 4}
\sum^{N}_{m=1}H(M_{m}\vert B)\geq -\log_{2}(b)+(N-1)S(A\vert B)+\max\{0 , \delta\},
\end{equation}
where  $$\delta=(N-1)I(A:B)-\sum^{N}_{m=1}I(M_{m}:B).$$In the case $\delta\geq0$, Eq.\;(\ref{Liu 4}) represents an improvement to Eq.\;(\ref{Liu 2}). It has been shown  that $\delta$ is a non-negative real number for many states such as the Bell diagonal states, the Werner states, and the maximally correlated mixed states \cite{Adabi}.
Zhang \emph{et al.}  introduced the entropic uncertainty relation for $N$ measurements $M_{m}$ as \cite{Zhang}
\begin{equation}\label{Zhang}
\sum^{N}_{m=1} H(M_{m})\geq(N-1)S(\rho)+\max_{u}\lbrace \ell^{U}_{u}\rbrace ,
\end{equation}
where
$$\ell^{U}_{u}=-\sum_{i_{N}}p_{u^{{N}}_{i_{N}}}\log_{2} \sum_{i_{k},N\geq k>1} \max_{i_{1}} \prod^{N-1}_{m=1}|\langle u^{m}_{i_{m}}|u^{m+1}_{i_{m+1}}\rangle|^{2},$$
and
$$p_{u^{{N}}_{i_{N}}}=Tr \big[ (|u^{N}_{i_{N}}\rangle\langle u^{N}_{i_{N}}| \otimes I )\rho_{AB} \big].$$
In the same way, one can convert this uncertainty relation to that in the presence of quantum memory,
\begin{equation}\label{zhang 2}
\sum^{N}_{m=1}H(M_{m}\vert B)\geq \max_{u}\lbrace \ell^{U}_{u} \rbrace+(N-1)S(A\vert B) +\max\{0 , \delta\}
\end{equation}
which is tighter than the Zhang uncertainty relation introduced in \cite{Zhang}. In another case, Xiao \emph{et al.} obtained the} following entropic uncertainty relation for multiple measurements \cite{Yunlong},
\begin{equation}\label{Xiao}
\sum^{N}_{m=1} H(M_{m})\geq(N-1)S(\rho)-\frac{1}{N}\omega B,
\end{equation}
{\bf where $ \omega $ indicates the universal majorization bound of $ N $ measurements and $ B $ is certain vector of logarithmic distributions.}
In the presence of quantum memory, Eq.\;(\ref{Xiao}) is converted to
\begin{equation}\label{Xiao 2}
\sum^{N}_{m=1}H(M_{m}\vert B)\geq (N-1)S(A\vert B)-\frac{1}{N}\omega B+\max\{0 , \delta\}.
\end{equation}
{\bf Based on what has been mentioned so far, one concludes that the method has two advantages. Firstly, it can be used to convert any entropy-based uncertainty relation in the absence of quantum memory to that in its presence. Secondly, the lower bounds of the uncertainty relations obtained by this method are tighter than those known so far.}

To illustrate these results, let us consider three observables $X=\sigma_x, Y=\sigma_y,$ and $ Z=\sigma_z$ measured on the part $A$ of the Werner state
 \begin{equation}\label{werner}
 \rho_{AB}=\eta \vert \psi_{+} \rangle_{AB}\langle \psi_{+} \vert + \frac{1-\eta}{4}\mathcal{I}_{AB},
 \end{equation}
where $\vert \psi_{+} \rangle_{AB} = 1/\sqrt{2}(\vert 0_A0_B \rangle + \vert 1_A1_B \rangle)$ is the maximally entangled state with $0 \leq \eta \leq 1$. {\bf The eigenvectors of the observables are given by} $$ X: \lbrace \frac{1}{\sqrt{2}}(1,1),  \frac{1}{\sqrt{2}}(-1,1) \rbrace,$$ $$Y: \lbrace \frac{1}{\sqrt{2}}( i,1),  \frac{1}{\sqrt{2}}(- i,1) \rbrace,$$ and $$Z: \lbrace ( 0,1), (1,0) \rbrace.$$
 \begin{figure}[t]
\centerline{\includegraphics[width=8cm]{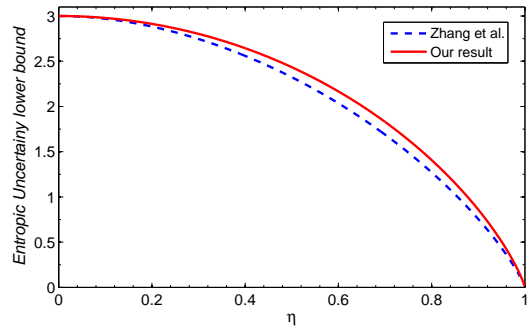}}
\vspace*{8pt}
\caption{(Color online) {\bf Variation of the lower bound of the entropic uncertainty relation of three complementary observables $X=\sigma_{x}, Y=\sigma_{y}$ and $Z=\sigma_{z}$ measured on the Werner state (Eq.\;(\ref{werner})) in the presence of quantum memory with respect to the parameter $\eta$. For comparison the plot of the Zhang entropic ULB is also included.}}\label{entropic}
\end{figure}
In Fig. \ref{entropic}, {\bf the lower bound of the entropic uncertainty relation for the measurement of three complementary observables $X=\sigma_{x}, Y=\sigma_{y}$ and $Z=\sigma_{z}$ measured on the Werner state is plotted versus the parameter $\eta$. As can be seen, the entropic ULB is tighter than that of Zhang.}

\section{Quantum coherence}\label{Sec4}
Quantum coherence represents an important feature of quantum physics that marks the deviation of quantum mechanics from the classical world. Regarding the definition of quantum coherence, the notion of quantumness is valid in single systems. Since coherence is defined to quantify the quantumness on a specified basis, it is logical to regard coherence as the generalized quantum
uncertainties \cite{Xiao Yuan,Xiao Yuan1}. {\bf Baumgratz \emph{et al.} introduced a  fundamental and accurate method to quantify coherence \cite{Baumgratz}.} To define a coherence measure, one has to know which states are incoherent. Consider  quantum states in a $d$-dimensional Hilbert space $\mathcal{H}_{d}$ with a particular basis $\lbrace \vert i \rangle \rbrace_{i=1,...,d}$.  All quantum states displayed by diagonal density matrix $\hat{\delta} =\sum_{i=1}^{d} \delta_{i}\vert i \rangle\langle i \vert $ in this particular basis are called incoherent states. This set of quantum states is represented by $\mathcal{I}$. Therefore $C$ can be considered as a proper coherence measure if it satisfies the following properties \cite{Baumgratz,Streltsov}
\begin{enumerate}
\item $C(\sigma) = 0$ if and only if $ \sigma \in \mathcal{I}$.
\item $C(\rho)$ is nonincreasing under incoherent completely positive trace preserving map (ICPTP) $\Lambda$,
i.e., $C(\rho) \geq C( \Lambda [\rho] )$. $\Lambda$ is an  ICPTP, if it can be written as $ \Lambda [\rho]=\sum_{m} K_{m} \rho K_{m}^{\dag}$, where $K_{m}$ 's are incoherent Kraus operators which map an incoherent state to  other incoherent state, and therefore $K_{m} \mathcal{I} K_{m}^{\dag} \subseteq \mathcal{I}$. This kind of map is known as non-selective ICPTP map.
\item $C(\rho)$ is nonincreasing under the selective ICPTP maps, i.e., $ C(\rho) \geq \sum_{m} p_{m}C(\rho_{m})$ in which $p_{m}=Tr[K_{m} \rho K_{m}^{\dag}]$ and $ \rho_{m}=K_{m} \rho K_{m}^{\dag} / p_{m}$.
\item $C(\rho)$ is a convex function for any set of states $\lbrace \rho_{i}\rbrace$ and any probability distribution $\lbrace p_{i}  \rbrace$.
\end{enumerate}
The $l_1$-norm of coherence and relative entropy of coherence \cite{Baumgratz} are proper coherence measures which satisfy the conditions listed above. {\bf The relative entropy of coherence is defined as
 \begin{equation}
C_{r}(\rho)=\min_{\delta \in \mathcal{I}} S(\rho\Vert \delta),
\end{equation}
where $S(\rho\Vert \delta)=Tr[\rho\log_{2}\rho]-Tr[\rho\log_{2}\delta]$ is the relative entropy.
It can be written in a simple form \cite{Baumgratz},
\begin{equation}
C_{r}(\rho)=S(\Delta(\rho))-S(\rho),
\end{equation}
where $\Delta(\rho)=\sum_{i} \langle i \vert\rho\vert i \rangle \vert i \rangle\langle i \vert$ indicates dephasing of
$\rho$ in the reference basis $\lbrace \vert i \rangle \rbrace$.

For a bipartite system $AB$, it is possible to consider the coherence with respect to a local basis on subsystem $A$. This is called unilateral coherence \cite{Teng Ma}. Thus, $A$-incoherent state (incoherent-quantum state \cite{Chitambar}) is defined with respect to the basis of subsystem $A$ ($\lbrace\vert i \rangle _{A}\rbrace$ ) as \cite{Jiajun Ma}
\begin{equation}
\sigma_{AB}\in \mathcal{I}_{B|A}, \quad \sigma_{AB}=\sum_{i}p_i \vert i\rangle\langle i \vert_{A}\otimes \sigma_{B|i},
\end{equation}
where $\mathcal{I}_{B|A}$ is the set of $A$-incoherent states, and $\sigma_{B|i}$ is arbitrary state of the subsystem $B$.} $A$-incoherent operation $\Lambda_{IC}^{B|A}$ takes $\mathcal{I}_{B|A}$ to itself. A measure of unilateral coherence for a bipartite density matrix $\rho_{AB}$ satisfies the abov-mentioned four properties and is given by
\begin{equation}\label{reou}
C_{r}^{B|A}=\min_{\sigma_{AB}\in \mathcal{I}_{B|A} }S(\rho_{AB}||\sigma_{AB}).
\end{equation}
{\bf Let us call it relative entropy of unilateral coherence. It can also be written as \cite{Chitambar}
\begin{equation}\label{reou2}
C_{r}^{B|A}=S(\rho_{AB}||\Delta_{A}(\rho_{AB}))=S(\Delta_{A}(\rho_{AB}))-S(\rho_{AB}),
\end{equation}
where $$\Delta_{A}(\rho_{AB})=\sum_{i}(\vert i\rangle\langle i \vert\otimes I)\rho_{AB}(\vert i\rangle\langle i \vert\otimes I),$$ is a local dephasing in subsystem $A$ in the chosen basis $\lbrace \vert i \rangle_{A} \rbrace$. It can also be written as
\begin{equation}\nonumber
\Delta_{A}(\rho_{AB})=\sum_{i}p_i \vert i\rangle\langle i \vert_{A}\otimes \rho_{B|i},
\end{equation}
in which $p_{i}= Tr_{AB} \big( (\vert i\rangle\langle i \vert\otimes I)\rho_{AB}(\vert i\rangle\langle i \vert\otimes I) \big),$ and $\rho_{B|i}= \frac{Tr_{A} \big( (\vert i\rangle\langle i \vert\otimes I)\rho_{AB}(\vert i\rangle\langle i \vert\otimes I) \big)}{p_{i}}$. }

\section{{\bf Uncertainty relation for quantum coherence}}\label{Sec5}
In this section it is shown that for bipartite quantum systems the unilateral coherence of a given quantum system in one measurement basis is restricted by that
of the same quantum system in other measurement basis. {\bf Furthermore, a lower bound on the sum of the unilateral coherences of a given state with respect to two different measurement bases is provided and it is shown that the lower bound is tighter than that derived from the Berta uncertainty relation. Using the uncertainty relation in the presence of quantum memory for multiple measurements, one obtains an uncertainty relation for unilateral coherence in several bases. For a bipartite system in the state $\rho_{AB}$, measuring one of two observables $X=\sigma_{x}$ and $Z=\sigma_{z}$ with eigenbases $\lbrace \vert x_{i} \rangle \rbrace$ and $\lbrace \vert z_{i} \rangle\rbrace$, respectively, on the part $A$, converts the state to}
\begin{figure}[ht]
\centering
\includegraphics[width=8cm]{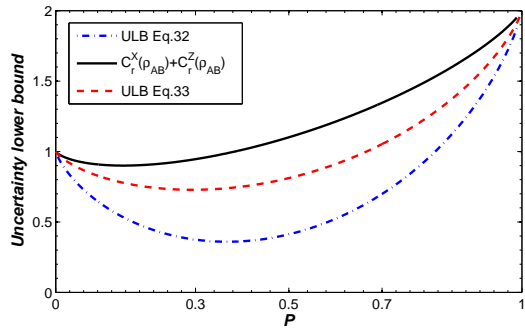}
\caption{(Color online) Lower bounds of the coherence-based uncertainty relation for two complementary observables $X=\sigma_{x}$ and $Z=\sigma_{z}$ in the presence of quantum memory, {\bf versus the parameter $p$ appeared in the expression for a correlated bipartite state which is assumed to be $\rho_{AB}= p|\Psi^{+}\rangle\langle\Psi^{+}|+(1-p)|11\rangle\langle11|$. The black (solid) line is $C_{r}^{X}(\rho_{AB})+C_{r}^{Z}(\rho_{AB})$, the red (dashed) and blue (dot dashed Line) curves represent  the ULB's in Eq.\;(\ref{new2}) and Eq.\;(\ref{bertacoherence}), respectively.}}\label{fig1}
\end{figure}
\begin{equation}
\rho_{QB}=\sum_{i} q_i \vert q_i \rangle\langle q_i \vert \otimes \rho_{B|q_i},
\end{equation}
{\bf in which $Q \in \lbrace X, Z \rbrace$ and $\vert q_{i} \rangle \in \lbrace \vert x_{i} \rangle,\vert z_{i} \rangle \rbrace$. $\rho_{QB}$ is the $A$-incoherent state with respect to $\lbrace\vert q_{i} \rangle\rbrace$ basis.} The trade-off relation between the measurement
uncertainty and its disturbance on the bipartite state $\rho_{AB}$ can be
written as \cite{Jun Zhang}
\begin{equation}\label{tor}
S(\rho_{AB}||\rho_{QB})=H(Q|B)-S(A|B).
\end{equation}
The term on the left hand side of {\bf Eq.\;(\ref{tor}) is in fact the relative entropy of unilateral coherence introduced in Eq.\;(\ref{reou2})} with respect to the measurement basis $\lbrace\vert q_{i} \rangle\rbrace$,
\begin{equation}\label{tor1}
C_{r}^{Q}(\rho_{AB})=H(Q|B)-S(A|B).
\end{equation}
Regarding Eq.\;(\ref{tor1}), for two incompatible quantum measurements (corresponding to two incompatible observables $X$ and $Z$) on the part $A$, one obtains
\begin{align}\label{new}
C_{r}^{X}(\rho_{AB})+C_{r}^{Z}(\rho_{AB})& =H(X|B)+ H(Z|B)-2 S(A|B).
\end{align}
{\bf This equation, along with Eq.\;(\ref{Berta}), leads us to
\begin{equation}\label{bertacoherence}
C_{r}^{X}(\rho_{AB})+C_{r}^{Z}(\rho_{AB}) \geq q_{MU}-S(A|B),
\end{equation}
which is a coherence-based uncertainty relation. Using Eq.\;(\ref{new1}), one finds another uncertainty relation which is
\begin{equation}\label{new2}
C_{r}^{X}(\rho_{AB})+C_{r}^{Z}(\rho_{AB}) \geq q_{MU} - S(A|B)+\max\{0 , \delta\}.
\end{equation}
In the case that $\delta\geq0$, the lower bound in Eq.\;(\ref{new2}) is tighter than that in Eq.\;(\ref{bertacoherence}).}

As an example, let us consider a special class of two-qubit $X$ states
 $$\rho_{AB}= p|\Psi^{+}\rangle\langle\Psi^{+}|+(1-p)|11\rangle\langle11|,$$
where $|\Psi^{+}\rangle=\frac{1}{\sqrt{2}}(|01\rangle+|10\rangle)$ is a maximally entangled state and $0\leq p\leq1$. In Fig. \ref{fig1}, {\bf the lower bounds of the coherence-based uncertainty relation for this state are plotted versus the parameter $p$. The plots show that the lower bound in Eq.\;(\ref{new2}) is tighter than that in Eq.\;(\ref{bertacoherence}).}
 \begin{figure}[t]
\centerline{\includegraphics[width=8cm]{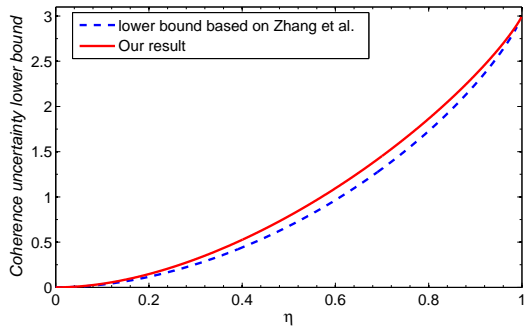}}
\vspace*{8pt}
\caption{(Color online) Lower bounds of the coherence uncertainty relation of three complementary observables $X=\sigma_{x}, Y=\sigma_{y},$ and $Z=\sigma_{z}$ in the presence of quantum memory, {\bf versus the parameter $\eta$. The state of the system is the Werner state.}}\label{coherence}
\end{figure}

Similarly, one can obtain the uncertainty relations for unilateral coherence in several bases using the relations obtained in Sec. \ref{Sec3}. For example, Eq.\;(\ref{zhang 2}) can be used to write an uncertainty relation for unilateral coherence in several bases,
\begin{equation}\label{zhang 5}
\sum^{N}_{m=1}C_{r}^{M_{m}}(\rho_{AB})\geq \max_{u}\lbrace \ell^{U}_{u}\rbrace -S(A\vert B)+\max\{0 , \delta\}.
\end{equation}
In Fig. \ref{coherence} {\bf the coherence ULB is plotted in terms of the parameter $\eta$. As can be seen, the lower bound in Eq.\;(\ref{zhang 5}) is tighter than that obtained by the Zhuang entropic uncertainty relation.}

\section{Conclusion}\label{conclusion}
The uncertainty relation has many applications in quantum information tasks, and many efforts have been made to expand and modify the relation. In this work, {\bf a method was introduced to transform the uncertainty relations in the absence of quantum memory into those in the presence of it. It was  also shown that the method provides tighter lower bound of uncertainty relation. Using the method, the uncertainty relations in the presence of quantum memory for multiple measurements were obtained. It was  also shown that the corresponding lower bounds are tighter than those of the similar uncertainty relations obtained previously through other methods. Regarding the uncertainty relations  in the presence of quantum memory, we could derive the  uncertainty relations for unilateral coherence. We also provided two uncertainty relations for the sum of unilateral coherence defined based on the relative entropy in two incompatible reference bases and compared the corresponding lower bounds. Using the uncertainty relation in the presence of the quantum memory for multiple measurements, uncertainty relation for unilateral coherence in several bases were obtained.}






\begin{thebibliography}{00}

\bibitem{Heisenberg} W. Heisenberg, Z. Phys. {\bf 43}, 172 (1927).
\bibitem{Robertson} H. P. Robertson, Phys. Rev. {\bf 34}, 163 (1929).
\bibitem{Schrodinger} E. Schr\"{o}dinger, Proc. Pruss. Acad. Sci. {\bf XIX}, 296 (1930).
\bibitem{Deutsch} D. Deutsch, Phys. Rev. Lett. {\bf 50}, 631 (1983).
\bibitem{Kraus} K. Kraus, Phys. Rev. D {\bf 35}, 3070 (1987).
\bibitem{Uffink} H. Maassen and J. B. M. Uffink, Phys. Rev. Lett. {\bf 60}, 1103
(1988).
\bibitem{Koashi} M. Koashi, New J. Phys. {\bf 11}, 045018 (2009).
\bibitem{Berta} M. Berta, M. Christandl, R. Colbeck, J. M. Renes, and R. Renner, Nat. Phys. {\bf 6}, 659 (2010).
\bibitem{Vallone} G. Vallone, D. G. Marangon, M. Tomasin, and P. Villoresi, Phys. Rev. A {\bf 90}, 052327 (2014).
\bibitem{Cao} Z. Cao, H. Zhou, X. Yuan, and X. Ma, Phys. Rev. X {\bf 6}, 011020 (2016).
\bibitem{Berta2} M. Berta, P. J. Coles, and S. Wehner, Phys. Rev. A {\bf 90}, 062127 (2014).
\bibitem{Walborn} S. P. Walborn, A. Salles, R. M. Gomes, F. Toscano, and P. H. Souto Ribeiro,
Phys. Rev. Lett. {\bf 106}, 130402 (2011).
\bibitem{Schneeloch} J. Schneeloch, C. J. Broadbent, S. P. Walborn, E. G. Cavalcanti, and J. C. Howell, Phys. Rev. A {\bf 87}, 062103 (2013)
\bibitem{Giovannetti} V. Giovannetti, S. Lloyd, and L. Maccone, Nature Photonics {\bf 5}, 222 (2011).
\bibitem{Coles1} P. J. Coles, M. Berta, M. Tomamichel, and S. Wehner, Rev. Mod. Phys. {\bf 89}, 015002 (2017).
\bibitem{Bialynicki} I. Bialynicki-Birula, Phys. Rev. A {\bf 74}, 052101 (2006).
\bibitem{Wehner} S. Wehner and A. Winter, New J. Phys. {\bf 12}, 025009 (2010).
\bibitem{Pati} A. K. Pati, M. M. Wilde, A. R. Usha Devi, A. K. Rajagopal and Sudha, Phys. Rev. A {\bf 86}, 042105 (2012).
\bibitem{Ballester} M. A. Ballester and S. Wehner, Phys. Rev. A {\bf 75}, 022319 (2007).
\bibitem{Vi}J. I de Vicente and J. Sanchez-Ruiz, Phys. Rev. A {\bf 77}, 042110 (2008).
\bibitem{Wu} S. Wu, S. Yu and K. Molmer, Phys. Rev. A {\bf 79}, 022104 (2009).
\bibitem{Rudnicki} L. Rudnicki, S. P. Walborn and F. Toscano, Phys. Rev. A {\bf 85}, 042115 (2012).
\bibitem{Pramanik} T. Pramanik, P. Chowdhury, and A. S. Majumdar, Phys. Rev. Lett. {\bf 110}, 020402 (2013).
\bibitem{Maccone} L. Maccone and A. K. Pati, Phys. Rev. Lett. {\bf 113}, 260401 (2014).
\bibitem{Coles} P. J. Coles and M. Piani, Phys. Rev. A {\bf 89}, 022112 (2014).
\bibitem{Adabi} F. Adabi, S. Salimi, and S. Haseli, Phys. Rev. A {\bf 93}, 062123 (2016).
\bibitem{Adabi1} F. Adabi, S. Haseli and S. Salimi, EPL  {\bf 115}, 6004 (2016).
\bibitem{Zozor} S. Zozor, G. M. Bosyk and M. Portesi,  J. Phys. A {\bf 47}, 495302 (2014).
\bibitem{R} L. Rudnicki, Z. Puchala and K. Zyczkowski, Phys. Rev. A {\bf 89}, 052115 (2014).
\bibitem{Liu} S. Liu, L.-Z. Mu, and H. Fan,  Phys. Rev. A {\bf 91}, 042133 (2015).
\bibitem{Zhang} J. Zhang, Y. Zhang, and C.-S. Yu, Sci. Rep. {\bf 5}, 11701 (2015).
\bibitem{Yunlong} Y. Xiao, N. Jing, S.-M. Fei, T. Li, X. Li-Jost, T. Ma, and Z.-X. Wang, Phys. Rev. A {\bf 93}, 042125 (2016).
\bibitem{Kamil} K. Korzekwa, M. Lostaglio, D. Jennings, and T. Rudolph, Phys. Rev. A {\bf 89}, 042122 (2014).
\bibitem{Rudnicki1} L. Rudnicki,  Phys. Rev. A {\bf 91}, 032123 (2015).
\bibitem{Pramanik1} T. Pramanik, S. Mal, and A. S. Majumdar, Quantum Inf. Process. {\bf 15}, 981 (2016).
\bibitem{Nielsen} M. L. Nielsen and I. L. Chuang, \textit{Quantum Computation and Quantum Information} (Cambridge University Press, Cambridge, 2000).
\bibitem{Xiao Yuan} X. Yuan, Q. Zhao, D. Girolami, and X. Ma, arXiv:1605.07818 (2016).
\bibitem{Xiao Yuan1} X. Yuan, G. Bai, T. Peng, and X. Ma, Phys. Rev. A {\bf 96}, 032313 (2017).
\bibitem{Baumgratz} T. Baumgratz, M. Cramer, and M. B. Plenio, Phys. Rev. Lett. {\bf 113}, 140401 (2014).
\bibitem{Streltsov} A. Streltsov, G. Adesso, and M. B. Plenio, Rev. Mod. Phys. {\bf 89}, 041003 (2017).
\bibitem{Teng Ma} T. Ma, M.-J. Zhao, H.-J. Zhang, S.-M. Fei, and G.-L. Long, Phys. Rev. A {\bf 95}, 042328 (2017).
\bibitem{Jiajun Ma} J. Ma, B. Yadin, D. Girolami, V. Vedral, and M. Gu, Phys. Rev. Lett. {\bf 116}, 160407 (2016).
\bibitem{Chitambar} E. Chitambar, A. Streltsov, S. Rana, M. N. Bera, G. Adesso, and M. Lewenstein, Phys. Rev. Lett. {\bf 116}, 070402 (2016).
\bibitem{Jun Zhang} J. Zhang, Y. Zhang, and C.-S. Yu, Int. J. Theor. Phys. {\bf 55}, 3943 (2016).


\end{thebibliography}




\end{document}